%% file: main.tex
\documentclass[%
reprint,
amsmath,amssymb,
aps,
twocolumn,
]{revtex4}  

\input{definitions.tex}

\begin{document}


\title{Explicit Granger Causality in Kernel Hilbert Spaces}


\author{Diego Bueso}
\author{Maria Piles}
\author{Gustau Camps-Valls}
\affiliation{Image Processing Laboratory (IPL), Universitat de Val\`encia, Spain.}



\begin{abstract}
Granger causality (GC) is undoubtedly the most widely used method to infer cause-effect relations from observational time series. 
Several nonlinear alternatives to GC have been proposed based on 
kernel methods.  
We generalize kernel Granger causality by considering the variables cross-relations explicitly in Hilbert spaces.  
The framework is shown to generalize the linear and kernel GC methods, and comes with tighter bounds of performance based on Rademacher complexity. 
We successfully evaluate its performance in standard dynamical systems, as well as to identify the arrow of time in coupled R\"ossler systems, and is exploited to disclose the El Ni\~no-Southern Oscillation (ENSO) phenomenon footprints on soil moisture globally.

\end{abstract}

\pacs{Valid PACS appear here}

\maketitle


\newpage

\section{\label{sec:level1}Introduction}

Establishing causal relations between random variables from observational data is perhaps the most important challenge in today's science, from  
Earth sciences~\cite{Runge19natcom} and neurosciences~\cite{reid2019advancing} to 
social sciences~\cite{marini1988causality}. 
Granger causality (GC)~\cite{granger_investigating_1969} was introduced as a first attempt to formalize quantitatively the causal relation between time series, and is the most widely used method. The intuition behind GC is to test whether the past of $X$ helps in predicting the future of $Y$ from its past alone. GC implicitly tells us about the concept of {\em information} using {\em forecasting}.
Other methods rely on similar concepts of {\em information flow} and {\em predictability}: connections can be established between GC and transfer entropy \cite{schreiber_measuring_2000}, directed information \cite{massey1990causality},
convergent cross-mapping \cite{sugihara_detecting_2012}, Liang's measure of information flow \cite{liang_information_2015},
and with the
graphical causal model perspective \cite{white_linking_2011}.

Noting the strong linearity assumption in GC~\cite{eichler_causal_2007}, nonlinear extensions of GC have been proposed, and many discussions on the validity of non-parametric test statistics and nonlinear GC models exist in the literature~\cite{cartwright2007hunting,diks2016nonlinear}. 
Several studies propose replacing the linear AR models with neural networks or random forests as forecasting methods: 
while improved  efficiency and model versatility are achieved, there is no principled statistical test to assess GC causality. 
A solid and mathematically sound approach comes from the field of kernel methods \cite{Rojo18dspkm}, which allows to develop nonlinear models from linear ones, while still resorting to linear algebra operations. 
Kernel methods have been widely used for regression, classification and dimensionality reduction. 
GC with kernels was originally introduced in \cite{Ancona}. The method assumed a particular class of functions and an additive interaction between them. An alternative kernel-based test in combination with a filtering approach  
was later introduced in \cite{marinazzo}. 
In all these studies, the autoregressive (AR) models use kernel-based regression on stacking the involved variables in input spaces. This approach, however, is limited as does not consider nonlinear cross-relations between $X$ and $Y$ explicitly in Hilbert spaces.  

We here introduce explicit feature maps and corresponding kernel functions that account for nonlinear cross-relations in kernel space \cite{martinez06}. 
We demonstrate that the cross-kernel methodology generalizes linear and kernel GC methods, come with statistical guarantees, and yield enhanced detection power.


\section{\label{sec:level2}Nonlinear Granger causality with kernels}
GC first builds univariate and bivariate AutoRegressive (AR) models: 
(1) $y_{t+1} =  \sum_{p=0}^P a_p y_{t-p} + \varepsilon_t^y$ and 
(2) $y_{t+1} = \sum_{p=0}^P a_p y_{t-p} + \sum_{q=0}^Q b_q x_{t-q} + \varepsilon_t^{y|x}$,
and then computes a GC test as the ratio of model fitting errors: $\delta_ {x \to y} = \log({\mathbb V}[\varepsilon_t^y]/{\mathbb V}[\varepsilon_t^{y|x}])$, where the residual errors are defined for the unrestricted $\varepsilon_t^y$ and restricted $\varepsilon_t^{y|x}$ cases separately, and ${\mathbb V}$ represents the variance operator. Time embeddings $P$ and $Q$ are selected by cross-validation or sensible  statistical criteria. 
Regressors are defined as $\y_t =[y_t,y_{t-1},\ldots,y_{t-P}]^\intercal$ and $\x_t= [x_t,x_{t-1},\ldots,x_{t-Q}]^\intercal$, and vector coefficients ${\bf a}=[a_1,\ldots,a_P]^\intercal$ and ${\bf b}=[b_1,\ldots,b_Q]^\intercal$ are typically estimated by least squares. 

\subsection{Feature maps and kernel functions}
The linear GC formulation can be generalized to the nonlinear case using elements of the theory of reproducing kernel Hilbert spaces (RKHS)~\cite{Rojo18dspkm}. Let us assume the existence of a Hilbert space $\mcalH$ equipped with an inner product 
where samples in $\mcalX$ are mapped into by means of a feature map $\bphi:\mcalX\to\mcalH, \x_i\mapsto \bphi(\x_i)$, $1\leq i\leq n$. The similarity between the elements in $\mcalH$ can be estimated using its associated dot product $\langle\cdot,\cdot\rangle_{\mcalH}$ via RKHS, $k:\mcalX\times\mcalX\to\mathbb{R}$, such that pairs of points $(\x,\x')$ $\mapsto$ $k(\x,\x')$. Therefore, one can estimate similarities in $\mcalH$ without the explicit definition of the feature map $\bphi$, and hence without even having access to the points in $\mcalH$.

An important concept in kernel methods is the representer or Riesz' representation theorem~\cite{RieNag55,Kimeldorf1971}. The {\em representer theorem} gives us the general form of the solution to the common loss formed by a cost (loss, energy) term and a regularization term.

In an RKHS ${\mathcal{H}}$, there exists a (kernel) function $k(\cdot,\cdot)$ such that $f(\x) = \sum_{i=1}^n\alpha_i k(\x,\x _i), \alpha_i\in\mathbb{R},  \balpha=[\alpha_1,\ldots,\alpha_n]^\intercal \in\real^{n}$, which is a linear combination of kernel functions. 
This property has been widely used to develop kernel methods for classification, clustering and regression~\cite{Rojo18dspkm}. 
Defining the regularized least squares functional, $L(y,\hat y) = \sum_{i=1}^n (y_i - f(\x_i))^2 + \lambda\|f\|_{\mathcal H}^2$ leads to the kernel ridge regression (KRR) method \cite{shawetaylor04,Rojo18dspkm}, which has a convenient analytic solution, $\boldsymbol{\alpha} = ({\bf K} + n\lambda {\bf I})^{-1}\y$, where $\lambda$ is the regularization term, ${\bf I}$ is the identity matrix, and ${\bf K}$ is the kernel matrix with entries $k(\x_i,\x_j)\in\real$. 
The KRR method is the preferred kernel method for nonlinear GC because of its simplicity (only one hyperparameter is involved) and good results in practice~\cite{marinazzo,Ancona}. 
In kernel GC, however, an important aspect has been largely disregarded: the proper definition of the mapping function that gives rise to the kernel function itself. Next, we formalize the field of kernel GC by proposing an explicit definition of the cross-terms between variables $X$ and $Y$ in Hilbert spaces.

\begin{figure}[t!]
\begin{center}
\setlength{\tabcolsep}{2.0pt}
\begin{tabular}{c c c}

(a) Stacked & (b) Summation & (c) Cross-kernel \\[2mm]

\begin{tikzpicture}[->,>=stealth',scale=0.8,transform shape,node distance=2cm,thin]
  \tikzstyle{everystate}=[fill=green!30,draw=none,text=white]
  
  \node[state] (xp) [fill=orange!30]{$x_{t-\tau}$};
  \node[state] (yp)   [below of = xp,fill=blue!30]{$y_{t-\tau}$};
  \node[state] (yt)  [right of=yp,xshift=0.4cm,fill=green!30] {$y_t$};

  \path[solid, bend right=320, thick](yp) edge node[xshift=-0.7cm,yshift=0.3cm] {\bf $\psi([x,y])$} (yt);
  \path[solid, bend left=360, thick](xp) edge node[xshift=-0.7cm,yshift=0.3cm] {\bf} (yt);
  
\end{tikzpicture}
&

\begin{tikzpicture}[->,>=stealth',scale=0.8,transform shape,node distance=2cm,thin]
  \tikzstyle{everystate}=[fill=green!30,draw=none,text=white]
  
  \node[state] (xp) [fill=orange!30]{$x_{t-\tau}$};
  \node[state] (yp)   [below of = xp,fill=blue!30]{$y_{t-\tau}$};
  \node[state] (yt)  [right of=yp,xshift=0.4cm,fill=green!30] {$y_t$};
  
  \path[solid, bend right=320, thick] (xp) edge node[xshift=0.5cm,yshift=0.3cm] {\bf $\psi(x)$} (yt);
  
  \path[solid, thick] (yp) edge node[xshift=0.0cm,yshift=-0.3cm] {\bf $\phi(y)$} (yt);
  
\end{tikzpicture}

&

\begin{tikzpicture}[->,>=stealth',scale=0.8,transform shape,node distance=2cm,thin]
  \tikzstyle{everystate}=[fill=green!30,draw=none,text=white]
  
  \node[state] (xp) [fill=orange!30]{$x_{t-\tau}$};
  \node[state] (yp)   [below of = xp,fill=blue!30]{$y_{t-\tau}$};
  \node[state] (yt)  [right of=yp,xshift=0.4cm,fill=green!30] {$y_t$};
  
  \path[solid, bend left=360, thick](xp) edge node[xshift=-0.7cm,yshift=0.3cm] {\bf} (yt);
  
  \path[solid, bend right=320, thick](yp) edge node[xshift=-0.7cm,yshift=0.3cm] {\bf $\widetilde\psi(x,y)$} (yt);

  \path[solid, bend right=320, thick] (xp) edge node[xshift=0.5cm,yshift=0.3cm] {\bf $\varphi(x)$} (yt);
  
  \path[solid, thick] (yp) edge node[xshift=0.0cm,yshift=-0.3cm] {\bf $\varphi(y)$} (yt);
  
\end{tikzpicture}

\end{tabular}
\end{center}
\vspace{-0.5cm}
\caption{
Representation of the GC model for the different kernel functions. Each model encodes explicit relations of past states with future ones. Note the difference between the regular map $\psi([\x,\y])$ working on the concatenation of time series in the input domain and the joint map $\widetilde \psi(\x,\y)$ working with the concatenation of maps of $\x$ and $\y$ in Hilbert spaces. 
\label{fig:graph}}
\end{figure}
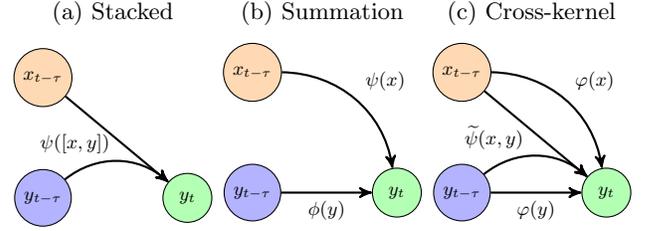

\noindent{\sf Stacked kernel.} 
The standard kernel GC (KGC) approach considers a straightforward approach to AR modeling with kernels~\cite{marinazzo,Ancona}, see Fig. \ref{fig:graph}(a). 
The method essentially defines two feature maps $\boldsymbol{\phi}$ and $\boldsymbol{\psi}$ to a RKHS ${\mathcal H}$ endorsed with reproducing kernels $k$ and $\ell$, where ${\bf y}_t$ and the concatenation $\z_t=[\y_t,\x_t]\in\real^{P+Q}$ are mapped to, respectively. This leads to the kernel regression models (1) $y_{t+1} = {\bf a}_H^T \bphi(\y_t) + \varepsilon_t^y$ and (2) $y_{t+1} = {\bf b}_H^T \bpsi(\z_t) + \varepsilon_t^{y|x}$, where now ${\bf a}_H,{\bf b}_H\in\real^{H\times 1}$. 
Now, by using the representer's theorems~\cite{RieNag55,Kimeldorf1971} on the model weights defined in RKHS, ${\bf a}_H =  \boldsymbol{\Phi}^\intercal \boldsymbol{\alpha}$ and ${\bf b}_H = \boldsymbol{\Psi}^\intercal \boldsymbol{\beta},$ where $\boldsymbol{\Phi},\boldsymbol{\Psi}\in\real^{n\times H}$, the AR models can be defined in terms of kernel functions only:
$y_{t+1} = \boldsymbol{\alpha}^\intercal{\bf k}_t + \varepsilon_t^y$, and $y_{t+1} = \boldsymbol{\beta}^\intercal\boldsymbol{\ell}_t + \varepsilon_t^{y|x}$, respectively, where ${\bf k}_t=[k(\y_1,\y_t),\ldots,k(\y_n,\y_t)]^\intercal$ and $\boldsymbol{\ell}_t=[\ell(\z_1,\z_t),\ldots,\ell(\z_n,\z_t)]^\intercal$ contain all evaluations of the kernel functions, $k$ and $\ell$ at time $t$, that act as similarity measures between the input feature vectors.
Importantly, note that since data are mapped to the same Hilbert space ${\mathcal H}$, the same kernel function and parameters are used for both $k$ and $\ell$. 


\noindent{\sf Summation kernel.} 
An alternative to the stacked approach builds {\em implicit} AR models in RKHS~\cite{Ancona} such that: $y_{t+1} = {\bf a}_H^T \bphi(\y_t) + \varepsilon_t^y$, and $y_{t+1} = {\bf a}_H^T \bphi(\y_t) +{\bf b}_H^T \bpsi(\x_t) + \varepsilon_t^{y|x},$
which leads to the kernel AR models 
$y_{t+1} = \alpha^\intercal{\bf k}_t + \varepsilon_t^y$ and $y_{t+1} = \alpha^\intercal{\bf k}_t + \beta^\intercal\boldsymbol{\ell}_t + \varepsilon_t^{y|x},$
where 
now $\boldsymbol{\ell}_t:=[\ell(\x_1,\x_t),\ldots,\ell(\x_n,\x_t)]^\intercal$. 
The summation kernel is more appropriate when large time embeddings $P$ and $Q$ are needed to capture long-term memory processes, since it avoids constructing large dimensional feature vectors $\z$ by concatenation, cf. Fig. \ref{fig:graph}(b). 
However, the cross-information between $X$ and $Y$ is missing 
\cite{martinez06}.

\noindent{\sf Explicit cross-kernel.} 
In order to account for cross-correlations in Hilbert space, 
we explicitly define two feature maps: the standard individual map $\phi$ and the joint feature mapping $\psi$ for the second AR model: $y_{t+1} = {\bf a}_H^T \boldsymbol{\phi}(\y_t) + \varepsilon_t^y$ and $y_{t+1} = {\bf b}_H^T \boldsymbol{\psi}(\x_t,\y_t) + \varepsilon_t^{y|x}$, 
where the joint map is defined by construction as
$\widetilde{\boldsymbol{\psi}}(\x_t,\y_t) := [{\bf A}_1\boldsymbol{\varphi}(\y_t),{\bf A}_2\boldsymbol{\varphi}(\x_t),{\bf A}_3(\boldsymbol{\varphi}(\y_t)+\boldsymbol{\varphi}(\x_t))]^\intercal,$ 
where $\boldsymbol{\varphi}$ is a nonlinear feature map into an RKHS ${\mathcal H}$, and ${\bf A}_i$, $i=1,2,3$, are three linear transformations from ${\mathcal H}$ to ${\mathcal H}_i$. The induced joint kernel function readily becomes:

\[
\begin{array}{ll}
&\hspace{-0.6cm}n((\x_t,\y_t),(\x_t',\y_t')) =\widetilde{\boldsymbol{\psi}}(\x_t,\y_t)^\intercal\widetilde{\boldsymbol{\psi}}(\x_t',\y_t') \\
 &=\boldsymbol{\varphi}(\y_t)^\intercal{\bf R}_1\boldsymbol{\varphi}(\y_t') + \boldsymbol{\varphi}(\x_t)^\intercal{\bf R}_2\boldsymbol{\varphi}(\x_t') \\ &~~+\boldsymbol{\varphi}(\y_t)^\intercal{\bf R}_3\boldsymbol{\varphi}(\x_t') + \boldsymbol{\varphi}(\x_t)^\intercal{\bf R}_3\boldsymbol{\varphi}(\y_t')\\
&=\!n_1(\y_t,\y_t')\!+\!n_2(\x_t,\x_t')\!+\!n_3(\y_t,\x_t')\!+\!n_4(\x_t,\y_t'),
\end{array}
\]

\noindent where ${\bf R}_1={\bf A}_1^\intercal {\bf A}_1+{\bf A}_3^\intercal {\bf A}_3$, 
${\bf R}_2={\bf A}_2^\intercal {\bf A}_2 + {\bf A}_3^\intercal {\bf A}_3$, 
and ${\bf R}_3={\bf A}_3^\intercal {\bf A}_3$. Note that the new kernel function considers cross-terms relations between the time series through kernels $n_3$ and $n_4$, and still works with the original time embeddings. Besides, there is no need to explicitly use the same kernel function or parameters. We now show that the cross-kernel GC (XKGC) method generalizes previous KGC methods and comes with statistical guarantees, see Fig.~\ref{fig:graph}(c).

\subsection{Statistical characterization with Rademacher complexity}\label{sec:rada}

Let us now characterize the generalization capabilities of the proposed cross-kernel using the notion of Rademacher complexity, which is perhaps the most useful measure used in the theoretical analysis and design of kernel algorithms~\cite{bartlett2002rademacher,koltchinskii2002empirical}. 
Rademacher complexity roughly states that one can infer (measure) an upper bound on the generalization performance of a given class by its ability to fit random data. The theory makes use of the Rademacher variables, and produces a measure of capacity called the Rademacher complexity. 

In what follows we give a bound of performance for the general case of compositions of kernels as in our proposed framework. 
Generalization bounds {based on Rademacher complexity} \cite{lanckriet2004learning,ying2009generalization} provide a strong theoretical foundation for a family of learning kernel algorithms based on convex combinations of base kernels, as in our case. 
Let us define a sample $S=\{({\bf x}_i,y_i)\}_{i=1}^n\in{\mathcal X}\times {\mathcal Y}$ generated by a distribution $D$ on a set ${\mathcal X}$, 
a family of functions ${\sf H}=\{h:{\mathcal X}\rightarrow \real\}$ 
and a loss function $L:{\mathcal X}\times{\mathcal X}\to\real^+$. The goal 
is to find the $h$ hypothesis in ${\sf H}$ with small generalization error with respect to the target $f({\bf x})$, $R_D(h)={\mathbb E}_{{\bf x}\sim D}[L(h({\bf x}),f({\bf x}))]$, empirically estimated as $R_D(h)=\frac{1}{n}\sum_{i=1}^n L(h({\bf x}_i),y_i))$, e.g. the least squares $L(y,\hat y)=(y-\hat y)^2$.

\noindent {\bf Theorem 1.} \emph{Generalization bound with finite ${\sf H}$.} Assuming a finite hypothesis set, ${\sf H}$ and that $L$ is bounded by $\varepsilon$, then for any $\delta>0$, with probability at least $1-\delta$,
$$R(h) \leq \widehat R(h) + \varepsilon\sqrt{\frac{\log|{\sf H}|+\log(2/\delta)}{2n}},$$
which can be particularized for the squared loss.


\noindent {\bf Theorem 2.} \emph{The least squares kernel regression bound.} Let $k:{\mathcal X}\times {\mathcal X}\to\real$ be a positive definite (PSD) kernel and $\boldsymbol{\phi}:{\mathcal X}\to{\mathcal H}$ be a feature map associated to $k$. Let the class of functions ${\sf H}=\{{\bf x}\mapsto{\bf w}^\top\boldsymbol{\phi}({\bf x}): \|{\bf w}\|\leq \Lambda\}$. Assume that both the kernel and the function are bounded, $k({\bf x},{\bf x})\leq M^2$ and $|f({\bf x})|\leq \Lambda M$ for all ${\bf x}\in{\mathcal X}$. The generalization risk $R$ is bounded by the empirical risk $\widehat R$ as follows. For $\delta>0$, with probability at least $1-\delta$ over random draws of samples of size $n$, in the sample $S$ for every $h\in {\sf H}$ satisfies:
$$R(h)\leq \widehat{R}(h)+\dfrac{8M^2\Lambda^2}{\sqrt{n}}\bigg(1+\dfrac{1}{2}\sqrt{\dfrac{\log(1/\delta)}{2}}\bigg),$$
and for every  $h\in {\sf H}$

$$R(h)\leq \widehat{R}(h)+\dfrac{8M^2\Lambda^2}{\sqrt{n}}\bigg(\sqrt{\dfrac{\text{Tr}[{\bf K}]}{nM^2}}+\dfrac{3}{4}\sqrt{\dfrac{\log(2/\delta)}{2}}\bigg).$$
This follows from the generalization bound with finite ${\sf H}$ and the bound on the Rademacher complexity of kernel hypotheses~\cite{bartlett2002rademacher,ying2009generalization}. Let us particularize this result for the cross-kernel and assess its generality.

\noindent{\bf Theorem 3.} 
\emph{Cross-kernel Rademacher complexity bounds.} 
The KRR function class $h$ uses the squared loss for $\widehat{R}(h)$. 
Let us assume a radial basis function (RBF) kernel, $k({\bf x}_i,{\bf x}_j)=\exp(-\|{\bf x}_i-{\bf x}_j\|^2/(2\sigma^2))$ so $k({\bf x},{\bf x})=1$, and let $\gamma\in[0,1]$ and $\beta\in[\gamma,1]$. The Rademacher complexity regression minimum bound for the cross-kernel is:
$$R_{\text{cross}}(h) \leq \widehat{R}(h) + \frac{8\|h\|^2}{\sqrt{n}}\left(\sqrt{\frac{1+\gamma}{1+\beta}}+\frac{3}{4}\sqrt{\frac{\log(2/\delta)}{2}}\right).$$
{\em Proof.} The Rademacher complexity for a sum of $N$ kernels $K_i$ can be easily bounded as $\widehat R(h) = \sqrt{N} \widehat R(h_i)$, $i=1,\ldots,N$. It is easy to see that $M^2=2(1+\beta)$ and the $\text{Tr}[{\bf K}]=2n(1+\gamma)$ for the cross-kernel. The result follows from substituting them in Theorem 1. Since for the stacked kernel, $M^2=1$ and $\text{Tr}[{\bf K}]=n$, and for the summation $M^2=2$ and $\text{Tr}[{\bf K}]=2n$, it follows that $R_{\text{cross}}(h)\leq R_{\text{sum}}(h) = R_{\text{stacked}}$.
Note that for $\gamma=\beta$, i.e. when $X$ and $Y$ convey correlated information, the cross-kernel bound converges to the stacked and the summation bounds. Interestingly, since $\gamma\leq \beta$, the cross-kernel bound will be always tighter than the stacked/summation bound, which are confirmed experimentally in Table \ref{tab:radamacher}.

\section{Experiments}\label{sec:results}
In all our experiments, we used the RBF kernel function and the KRR method. Hyperparameters (regularization term $\lambda$ and kernel lengthscale $\sigma$) were selected by cross-validation. The statistical test of robustness was computed as in KGC~\cite{marinazzo}, and the threshold was set to the highest causal strength estimated from $100$ surrogate time series \cite{quiroga2002}. For the sake of reproducibility, code snippets and demos are provided in \href{http://isp.uv.es/code/xkgc.zip}{XKGC}. We compare GC, KGC (stacked and summation kernels are theoretically identical, and deemed similar in practice in low-dimensional settings) and XKGC in all experiments.

\subsection{Nonlinear coupled system}
\begin{figure}[h!]
\begin{center}
\setlength{\tabcolsep}{2pt}
\centerline{
\rotatebox{90}{\hspace{0.9cm} $Frequency$} ~~
\hspace{-0.3cm}\includegraphics[width=8.5cm]{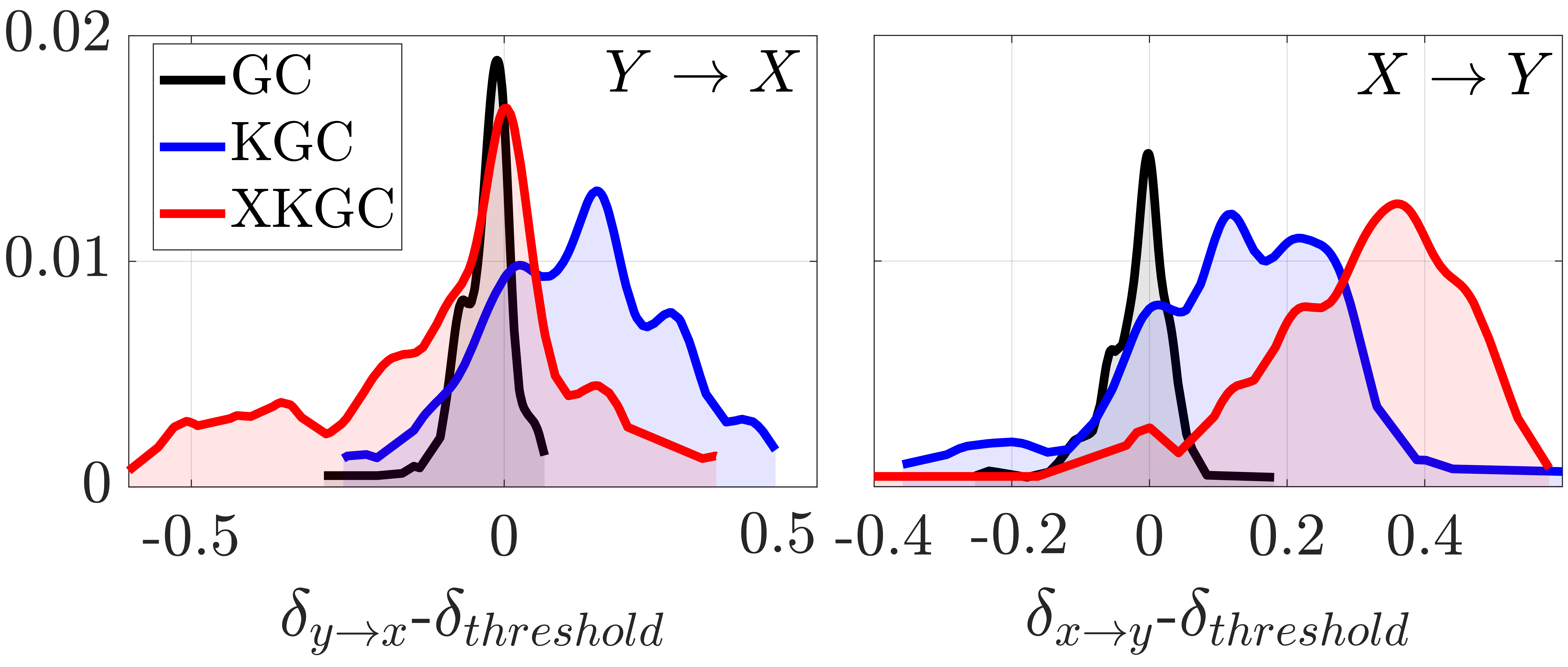}}
\end{center}
\vspace{-1cm}
\caption{Significance of the positive and negative cases detected for the coupled AR system. Histogram of the difference between estimated causality index $\delta$ and associated threshold $\delta_{threshold}$ are shown for each method and direction. 
\label{fig:nonlin}}
\end{figure}

Let us first consider a bivariate system with strongly coupled, non-linear and autoregressive relations defined as $x_{t+1} = 3.4x_{t}(1-x_{t}^2)\exp(-x_{t}^2)+\varepsilon_{t}^x$ and $y_{t+1} = 3.4y_{t}(1-y_{t}^2)\exp(-y_{t}^2)+\frac{x_{t}y_{t}}{2}+\varepsilon_{t}^y$, where $\varepsilon$ is white Gaussian noise with zero mean and variance 0.4. The causal direction is $x \to y$, being the opposite direction anti-causal. Standard GC, KGC and XKGC were run on a set of $n=4000$ samples 
and repeated $10,000$ times.
Figure~\ref{fig:nonlin} shows the histogram of the estimated causality index. Results reveals the insensitivity of linear GC to the causal direction and the high false positive rate of KGC, while XKGC shows a higher detection power and lower rates of false positives and true negatives.

\subsection{Logistic maps}

\begin{figure}[h!]
\centerline{\includegraphics[width=8.5cm]{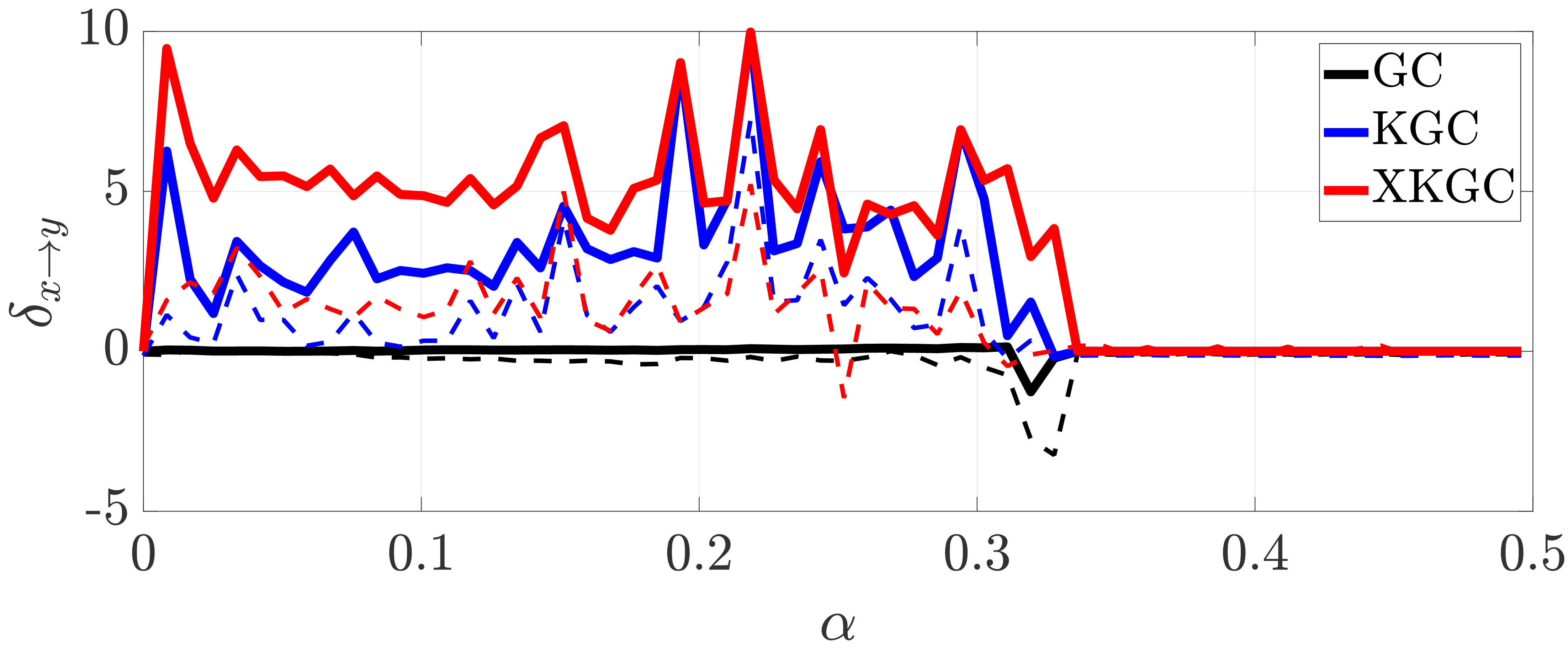}}
\vspace{-0.25cm}
\vspace{-0.25cm}\caption{Causality index $\delta$ estimated for each method (solid) and their associated thresholds (dashed) for the coupled logistic maps as a function of the coupling parameter $\alpha$.  }\label{fig:logistic}
\end{figure}

The second example considers the standard system of two logistic maps, defined as $x_{t+1} = 1 - 1.8x_{t}^2$ and $y_{t+1} =  (1-\alpha)(1 - 1.8y_{t}^2) +\alpha(1 - 1.8x_{t}^2)$,  where $\alpha \in [0,1]$ controls the coupling strength. The causal relationship implemented is $X\to Y$, and the challenge is to assess the detection power of methods without introducing any external variable, just using $X$ and $Y$. We analyze segments of length $n=2000$ and fixed $p=2$.  Figure~\ref{fig:logistic} shows the prediction skills for varying $\alpha$. Note that the system is completely synchronized at $\alpha=0.37$. The XKGC method shows improved detection power in the whole solution range.

We confirmed empirically the theoretical results in \S\ref{sec:rada} for the system of two logistic maps with $\alpha$=0.1. Empirical results of the Rademacher complexity bounds are provided in Table \ref{tab:radamacher}, where the cross-kernel achieves tighter bounds.

\begin{table}[h!]
\caption {Complexity terms for logistic maps.} \label{tab:radamacher} 
\vspace{-0.25cm}
\begin{center}
\begin{tabular}{ |P{1.5cm}||P{2.1cm}|P{2.1cm}|P{2.1cm}|  }
\hline
& {\bf Stacked} & {\bf Summation} & {\bf Cross-kernel} \\
\hline
$R$       & 0.4390  &  0.4361 & ${\bf 0.4352}$\\
\hline
$\widehat R$    & 0.6283 &  0.5001  & ${\bf 0.4992}$\\
\hline
\end{tabular}
\end{center}
\end{table}

\subsection{The arrow of time}

Let us now exemplify the performance of the proposed methods in the challenging problem of detecting the arrow of time from bivariate time series. This is a mostly academic question that has captured the attention in the physics literature \cite{reichenbach1991direction}, where both theoretical \cite{bauer2016arrow} and experimental \cite{Palus2018} results recently confirmed identifiability. 

We study the coupled R\"{o}ssler system which encompasses the prediction of the causal direction between two variables as well as the identification of the direction of time. The R\"ossler systems were originally introduced in the 1970s as prototype equations for the study of continuous-time chaos. The bivariate system studied in our paper was extracted from \cite{palus2007}. 
The unidirectional bivariate coupled R\"{o}ssler system was tested for a coupling parameter of $\epsilon = 0.07$, delay parameter $\delta=0$ and 6000 samples. Characteristic parameters for each system and initial conditions remain as in the original work.
We estimated detection power in both forward and backward propagation by just flipping the time series. 

\begin{figure}[t!]
\begin{center}
\setlength{\tabcolsep}{0.0pt}
\begin{tabular}{ccccc}
 & \multicolumn{2}{c}{Forward} & \multicolumn{2}{c}{Backward} \\
\rotatebox{90}{\hspace{1.2cm} $\delta$}
&
\includegraphics[height=2.4cm]{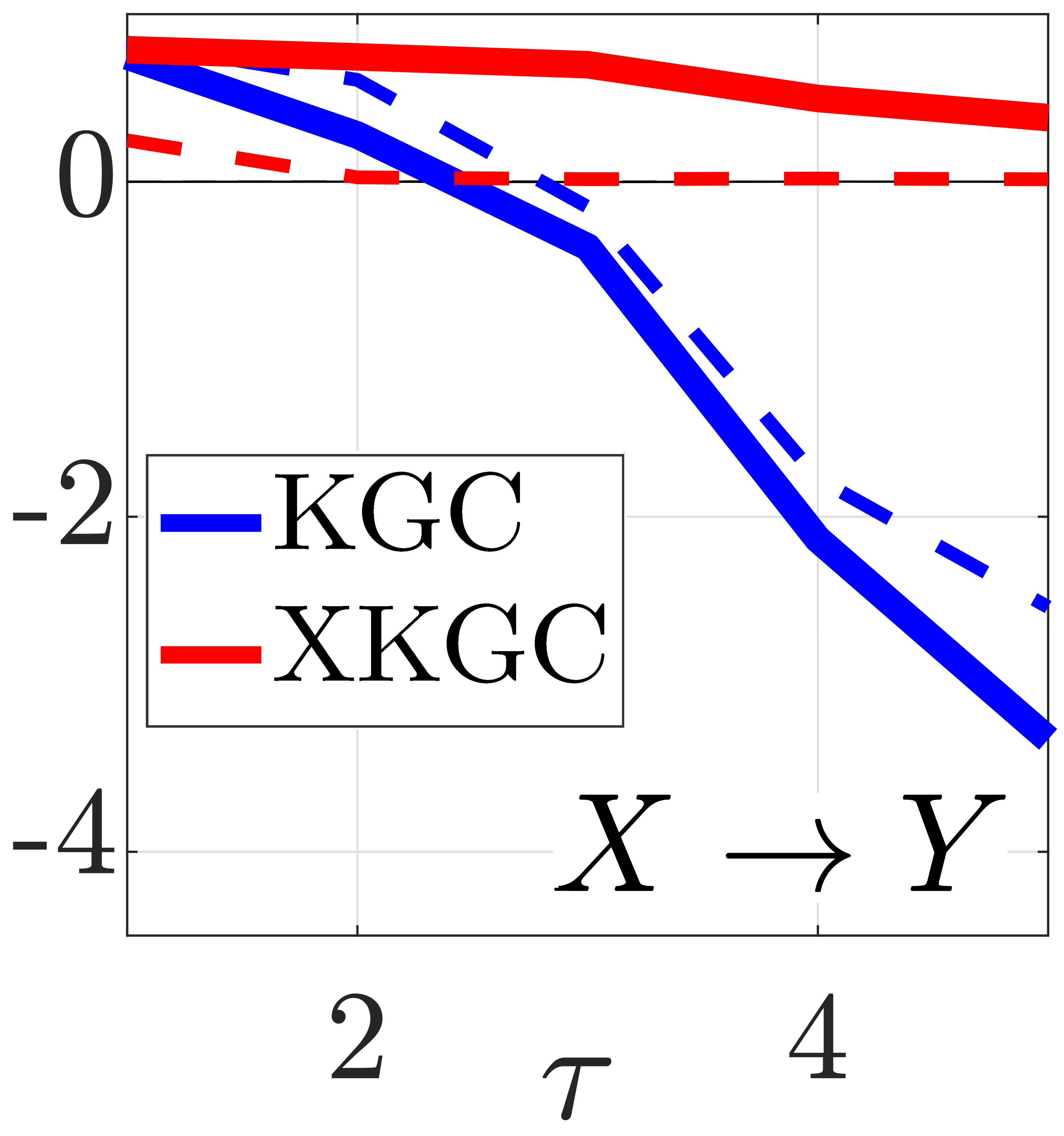}
&
\includegraphics[height=2.39cm]{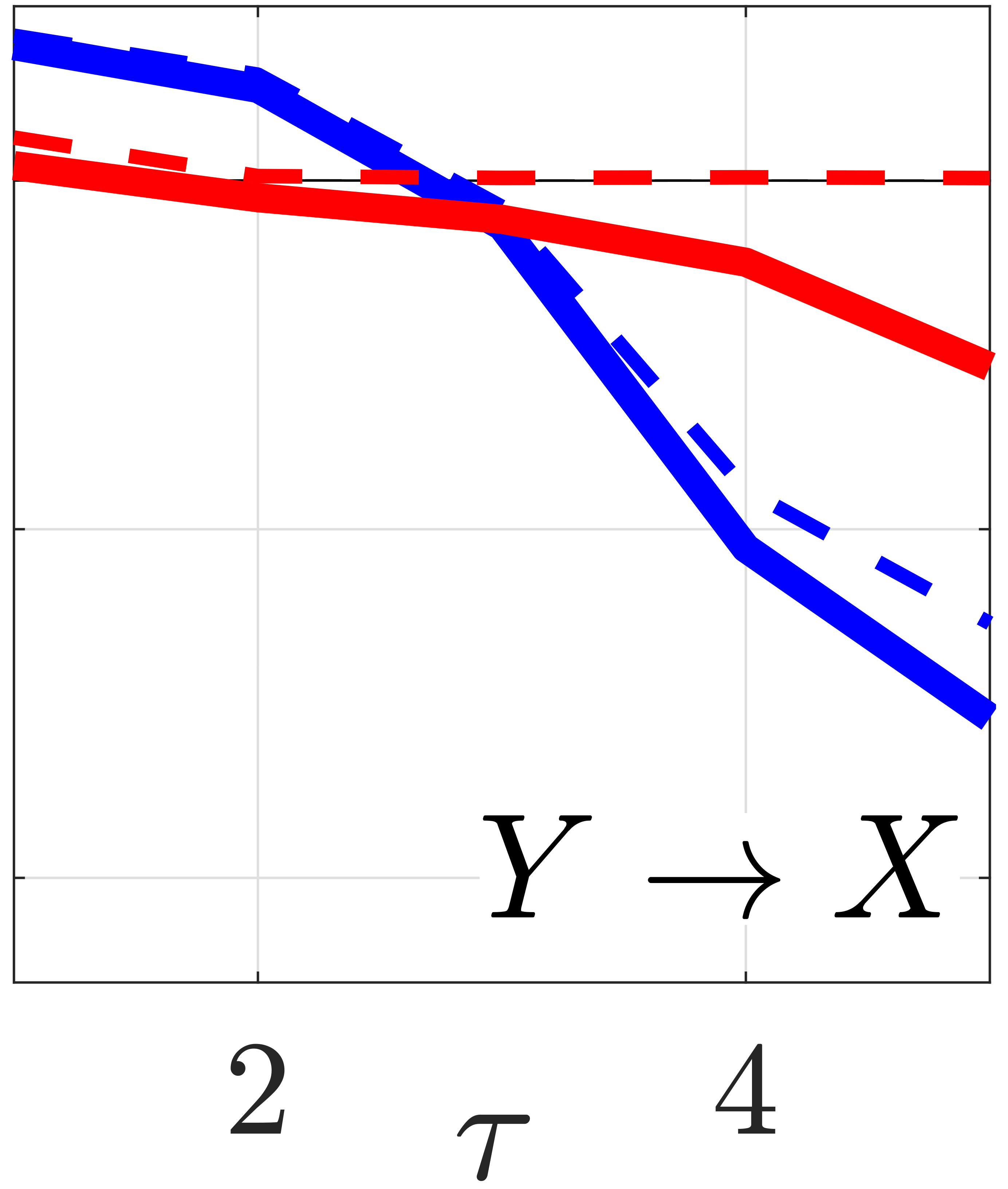}
&
\includegraphics[height=2.39cm]{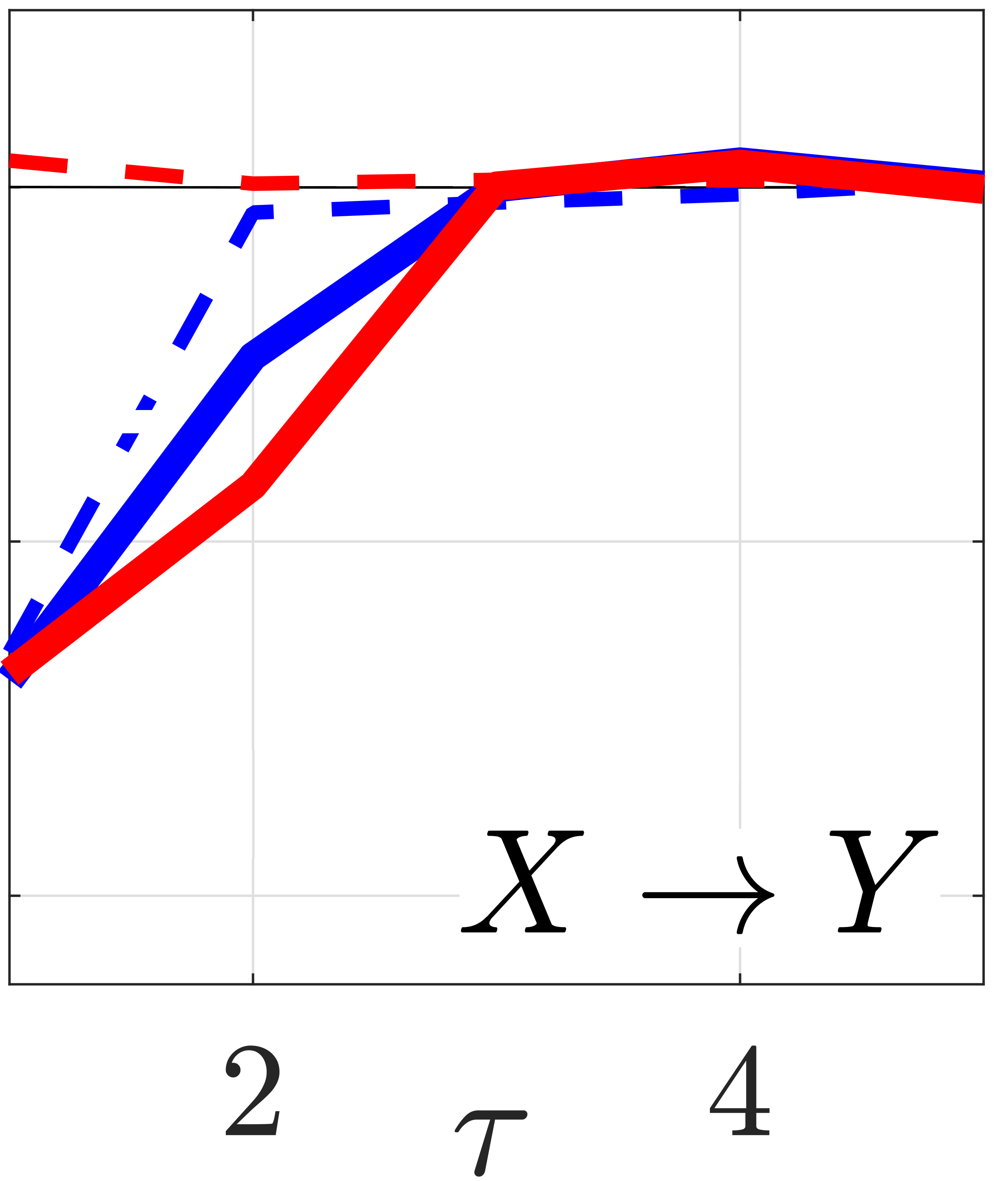} 
&
\includegraphics[height=2.39cm]{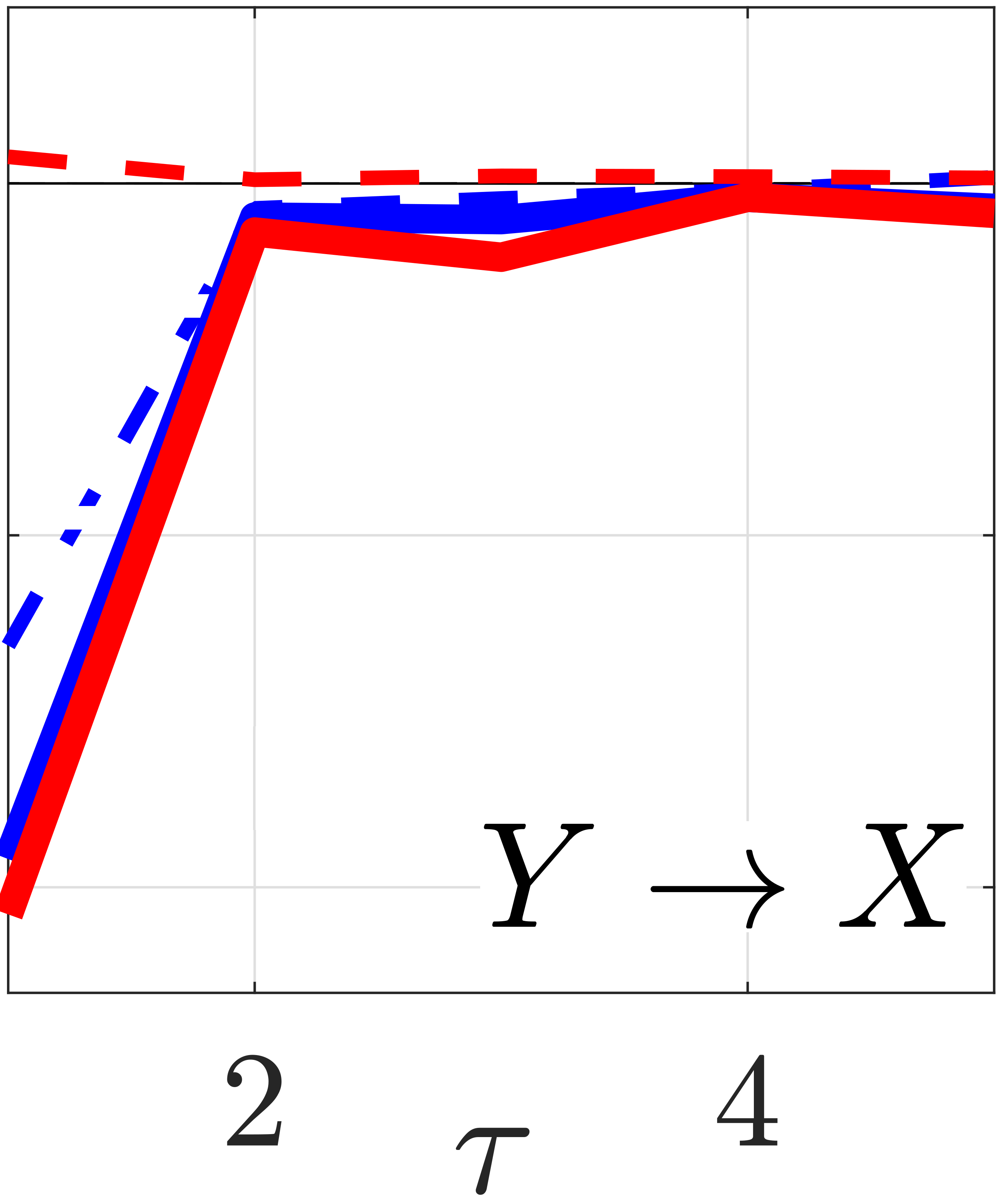}\\
\end{tabular}
\end{center}
\vspace{-0.5cm}
\caption{Arrow of time and the causal direction for a coupled R\"{o}ssler system. Causal strength estimated for each method (solid) and associated thresholds (dashed) for forward and backward propagation directions against the time delay $\tau$. \label{fig:AoT}}
\vspace{-0.25cm}
\end{figure}

Figure~\ref{fig:AoT} shows that the physical nature of the coupling system emerges as a forward propagation and with $X\to Y$. XKGC is the only method that can reconstruct the causal direction over the time delay order $\tau$ properly. This suggests that the proposal captures an extra variability of the coupled system, which in turn helps causal inference.

\begin{figure*}[ht!]
\begin{center}
\setlength{\tabcolsep}{2pt}
\begin{tabular}{c}
\includegraphics[width=8cm]{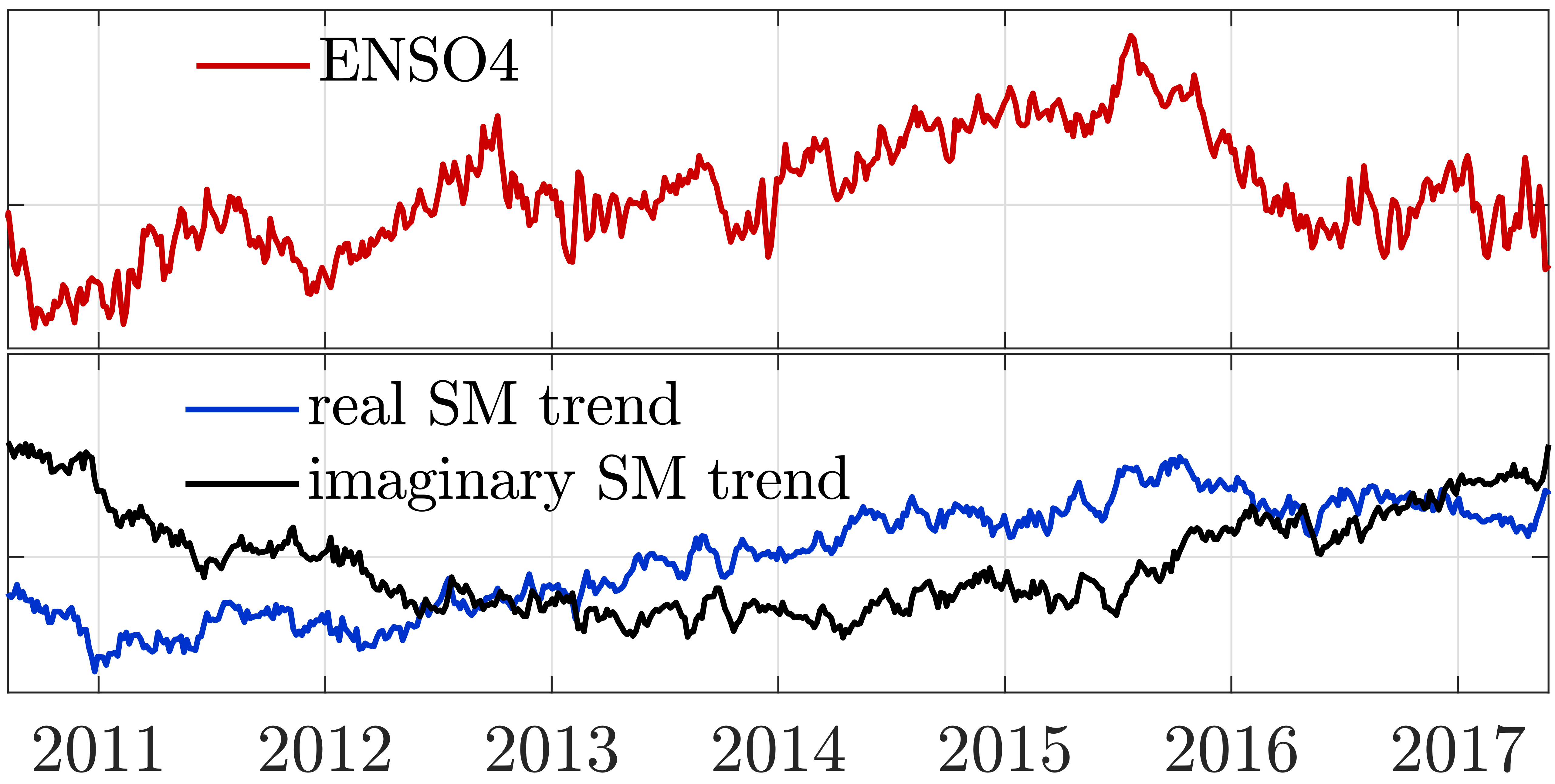} \\[2pt]
\includegraphics[width=8cm]{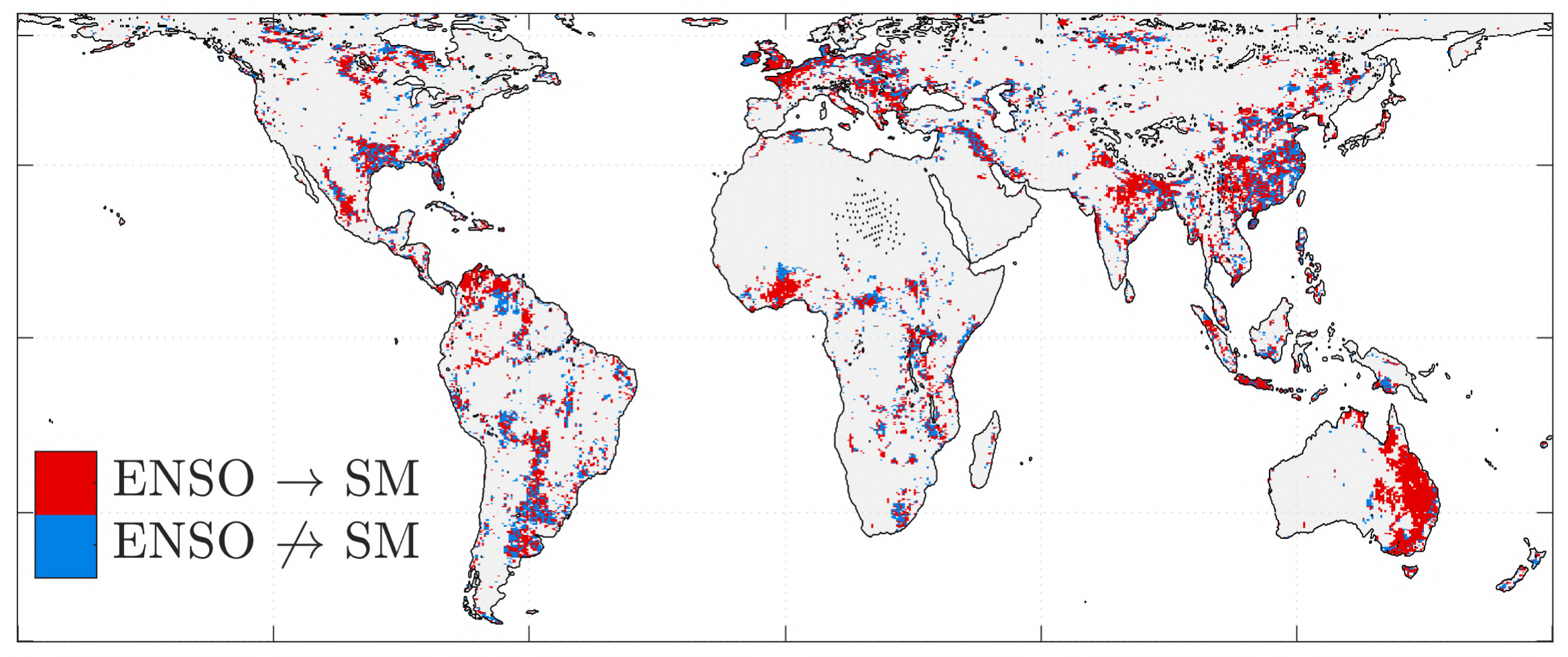}\\
\end{tabular}
\begin{tabular}{ccc}
\rotatebox{90}{\hspace{0.6cm} (a) GC} &
\includegraphics[width=0.68\columnwidth]{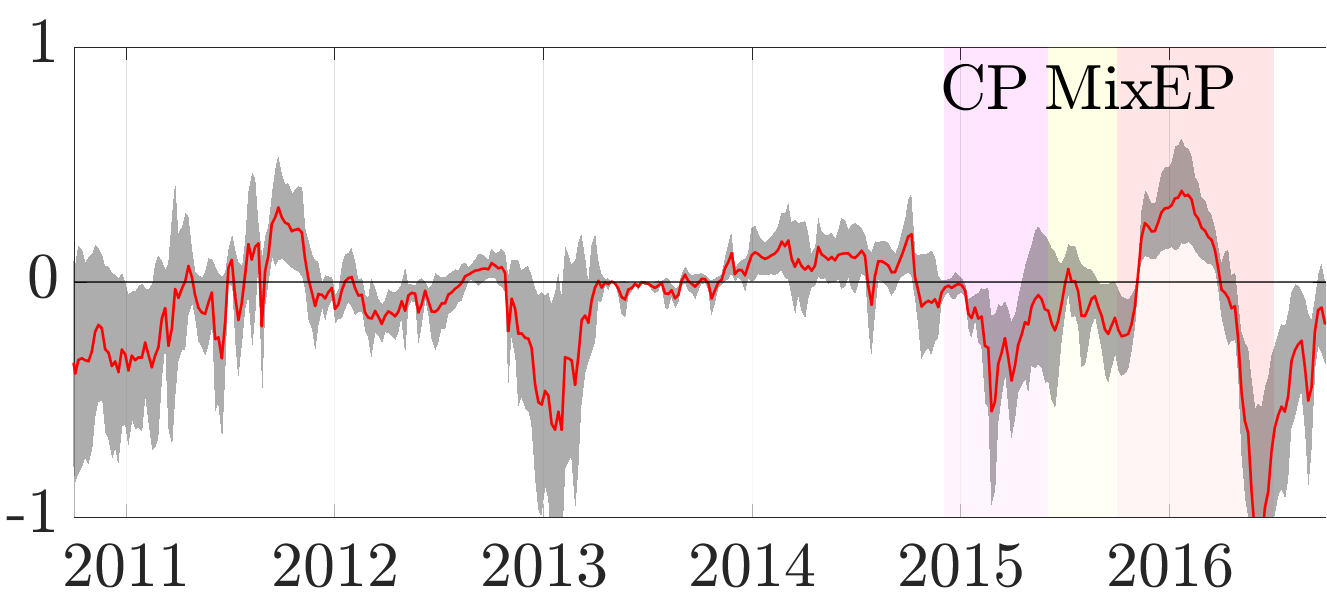} &
\includegraphics[width=0.34\columnwidth]{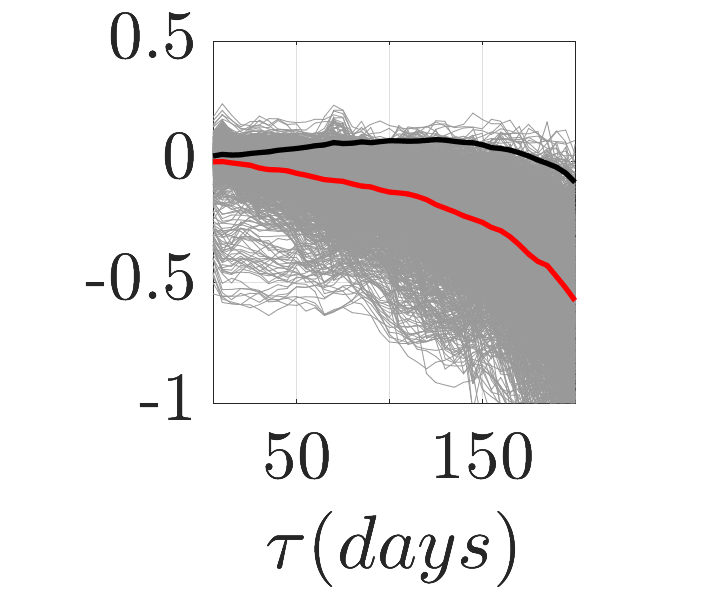} \\

\rotatebox{90}{\hspace{0.6cm} (b) KGC} &
\includegraphics[width=0.68\columnwidth]{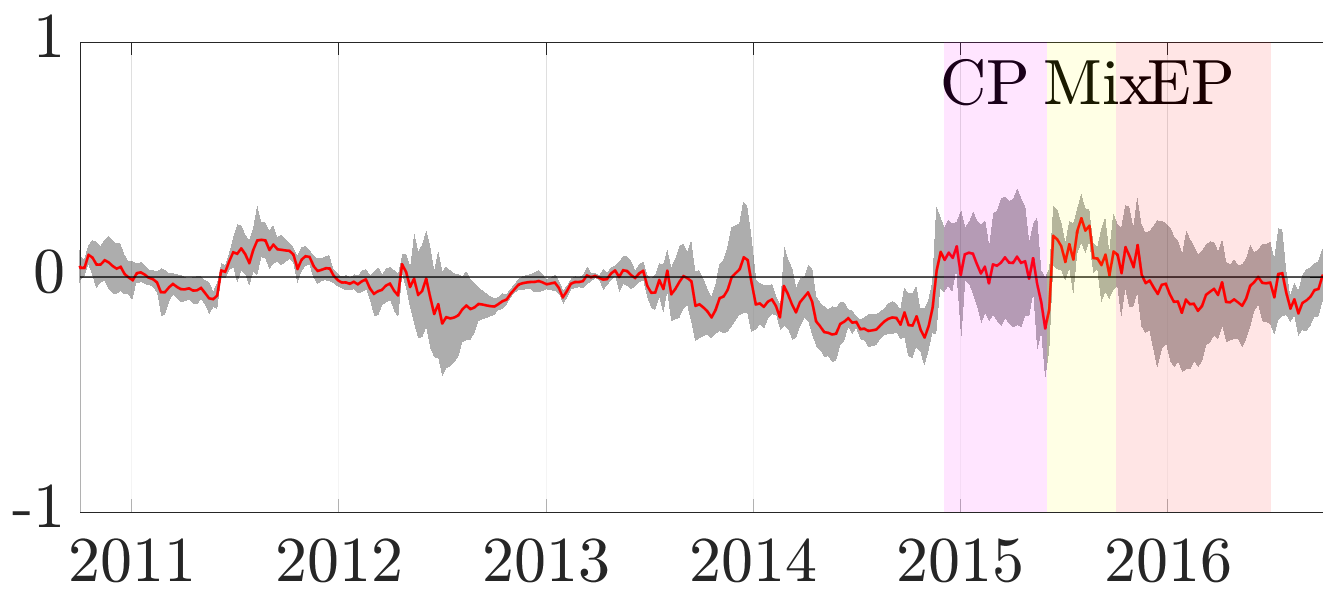} &
\includegraphics[width=0.34\columnwidth]{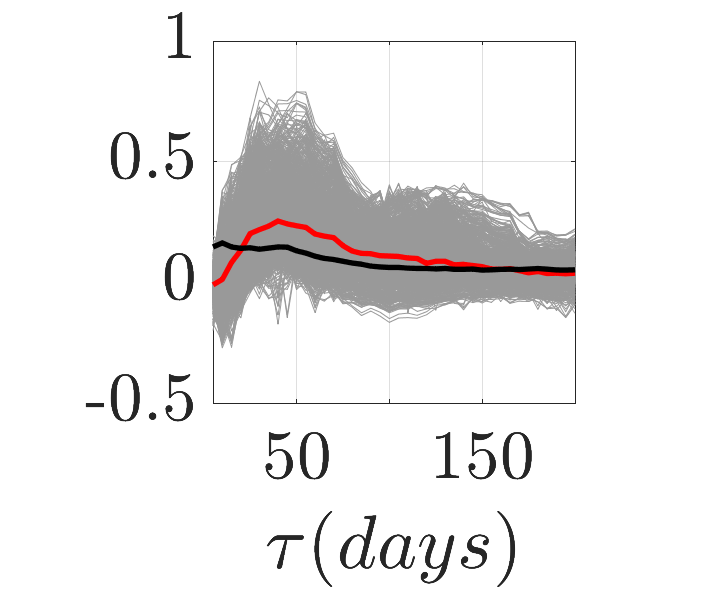} \\

\rotatebox{90}{\hspace{0.6cm} (c) XKGC} &
\includegraphics[width=0.68\columnwidth]{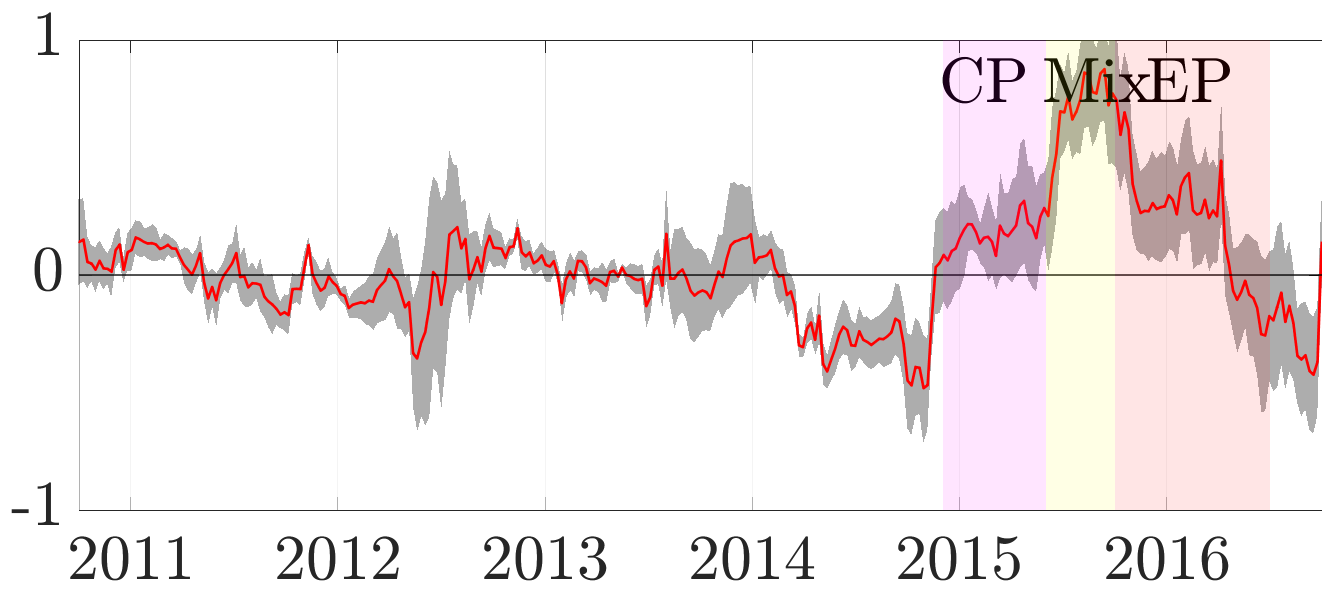} &
\includegraphics[width=0.34\columnwidth]{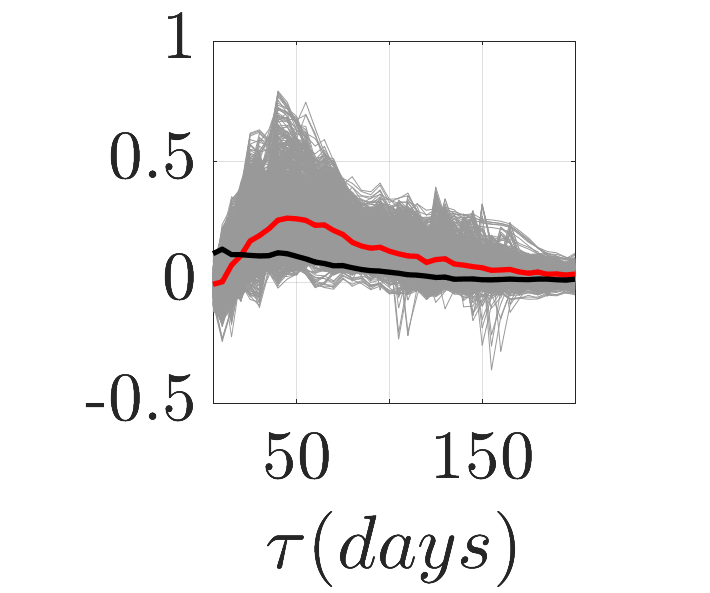} \\
\end{tabular}
\end{center}
\vspace{-0.5cm}
\caption{
{\em Top left:} ENSO4 (red) and estimated SM inter-annual complex components (blue and black) studied.
{\em Right panel:} GC analysis of ENSO and SM. Left plots show  $\delta$ (red line) with GC (a), KGC (b) and XKGC (c) estimated with a 2-year moving window, its variability (shaded), and highlights the three phases in the 2016 ENSO event (CP: Central Pacific, EP: East Pacific, Mix: Mixture of both cases). Right plots show the distribution (gray) and averages (red) of $\delta$ and thresholds (black) for lag $\tau$ and $100$ runs.
{\em Bottom left:} Spatial distribution of causal footprints of ENSO on SM obtained by XKGC.
} \label{fig:ensosm}
\end{figure*}

\section{Causal footprints of El Ni\~{n}o on soil moisture}
Causal discovery in Earth and climate sciences is a fundamental topic of research, as it allows systematic hypothesis testing, model-data inter-comparison, and discovery of patterns and causal links in observational data~\cite{Runge19natcom}. The challenges are multifaceted: Earth data shows spatio-temporal dimensions, complex nonlinear dynamics and teleconnections. We here tackle the problem of inferring causal links between El Ni\~{n}o/Southern Oscillation (ENSO) and soil moisture (SM) globally. ENSO is a coupled ocean-atmosphere phenomenon, which manifests as a quasi-periodic fluctuation in sea-surface temperature and air pressure in the equatorial Pacific Ocean. Although the exact causes initiating warm or cool ENSO events are not fully understood, the two components of ENSO --sea-surface temperature and atmospheric pressure-- are strongly related \cite{neelin98}. During the ENSO event, the atmospheric current of equatorial Walker circulation stops and changes its westward propagation for a more eastward direction. This transition occurs between 3-4 months before the rise of the sea surface temperature and is related with the ocean kelvin waves delay \cite{yang2013}. This disruption in the normal ocean-atmosphere coupling affects the propagation of the low pressure centers over the tropical regions, influencing temperature and precipitation across the globe. ENSO is hence strongly connected with global dry/wet anomalies, mostly over the tropics \cite{trenberth98}, but also over supra-tropical regions \cite{alexander02}.
ENSO and SM are connected by the atmospheric current and their causal relation is interrupted by its variations. 

Our causal analysis aims to uncover (spatially explicit) dry and wet patterns and identify footprints of ENSO on SM using satellite-based measurements. We use global SM maps from the \href{http://bec.icm.csic.es/}{ESA SMOS mission} and time series of the
\href{https://www.knmi.nl/home}{ENSO4 climate index} for the period 2010-2017. 
The information for the dataset we use here is introduced in Appendix \label{appendixA}. Our analysis focuses on the 2015-2016 ENSO event which had a strong impact over the atmosphere circulation \cite{Newman2016},  
being one of the three strongest El Ni\~{n}o events on record and the one with longest duration \cite{Zhai2016}. 
ENSO events can emerge with different spatial patterns (called flavours as well) \cite{Johnson2013}, which dominate the evolution of the ocean-atmosphere feedbacks and therefore the teleconnection patterns \cite{wiedermann2017}. 

Previous studies have shown that inter-annual variability in SM reflected known ENSO teleconnection patterns \cite{miralles2013,Piles19RS}. However, these works only focused on association (correlation), not on predictability (Granger causality). In this work, we are concerned about two important questions of the ENSO-SM coupled system: 
1) can we identify the different phases of the ENSO event and its transitions from the neutral state from purely observational data?, and
2) what part of the SM inter-annual signal is Granger caused by ENSO?
Answering these two questions allows us to revisit the ENSO-SM teleconnection map, to find unreported footprints of ENSO on SM globally.

The information in the global datacubes of SM need to be summarized in the so-called `modes of variability', i.e. spatially and temporally intrinsic components describing regional subprocess interactions. This is typically done with dimensionality reduction methods, such as PCA/EOF. PCA can only achieve a {\em linear} and {\em orthogonal} feature representation, which are not appropriate to deal with the highly nonlinear and interdependent nature of Earth observational data. As an alternative, a complex-valued non-linear PCA analysis was applied here to extract the dominant modes of global SM variability for the study period \cite{Bueso20rock}. 
The method allows us to extract nonlinear features that have independent spatial and temporal components into the complex domain.  The so-called ROCK-PCA method performs the eigendecomposition of a kernel matrix using data in the complex domain after applying the Hilbert transform~\cite{HilbertTransform}, and further rotated with a Promax transform~\cite{promax}. Complex-valued processes return us more useful components, as for example, the interpretation of phase-modulation decomposition against only the real part returned by regular PCA. In addition, the nonlinear nature of ROCK-PCA allows us to better capture feature relations. We provide source code of the method in \href{https://github.com/DiegoBueso/ROCK-PCA}{ROCK-PCA} \cite{Bueso20rock}.

In our experiments, we focus on the extracted inter-annual component of the global soil moisture satellite data, which represents 10.2\% of the total variance. We will use the temporal feature to estimate the link with ENSO and its spatial representation to identify the regions where the mode is relevant, i.e. its spatial amplitude is greater than the median of the spatial amplitudes with one positive standard deviation. The inter-annual SM component is lag-correlated (80 days) with ENSO4, $\rho \sim 0.8$ and co-integrated, see Fig.~\ref{fig:ensosm}(a). A map of the ENSO-SM causality index and details on its spatialization are provided in Appendix B. 

We analyzed the causal relation over a 2-year moving window to deal with non-stationarity~\cite{Stuecker} and studied model's $\delta$ sensitivity following \cite{nicolau2016}. 
We studied the model's sensitivity by jittering parameters for each trained model resulting from a different combination of window and time embedding~\cite{nicolau2016}. The model was also trained for several time embeddings to find the optimum time delay of the variability shared between the signals. For each trained model (window and time embeding) we have also estimate the sensibility of the causal index introducing a slightly perturbation of the model parameters \cite{nicolau2016}.
Results in Fig.~\ref{fig:ensosm}[right panel] reveal clear differences between the linear and the non-linear  $\delta_{\text{ENSO}\to\text{SM}}$ both across time and per time embedding. KGC and XKGC yield similar results yet differ in the magnitude of the  captured variability. Note that XKGC more clearly differentiates the three phases and the atmospheric disruption before ENSO rises. 
The impact of ENSO on the spatial distribution of the SM inter-annual trend is analyzed in Fig.~\ref{fig:ensosm}[bottom left]. On average, XKGC results indicate that about 50\% of SM interannual variability is caused by ENSO. Regions where SM interannual variability can be predicted by ENSO reproduce the well-known ENSO-induced precipitation patterns and teleconnections~\cite{Dai2000,Lyon2005,Yeh2018,Kim2013}, 
While some regions are clearly dominated by ENSO (e.g. Australia), others can only partially be explained by it (e.g. Gulf of Mexico, SE Asia), probably due to the influence of other atmospheric hydroclimatic patterns such as the Madden-Julian Oscillation \cite{Tang2008}. Notably, XKGC uncovers the impact of ENSO in yet unreported areas (e.g. NW Europe). Investigating additional causes dominating SM inter-annual variability and the emergence of potential new teleconnection patterns is recommended for future research.

\section{Conclusions}
We considered the problem of Granger causality and proposed a kernel-based framework that generalizes linear GC and KGC approaches. 
The theory of reproducing kernel functions allows us to derive different nonlinear algorithms while still resorting to linear algebra operations. 
The methodology copes with nonlinear relationships more efficiently and comes with statistical guarantees.

The methodology outperformed linear and nonlinear counterparts in 
standard dynamical systems, the arrow of time problem, and a real Earth system science problem.  
We expect that the generalized kernel Granger causality framework introduced here paves the way to enhanced models through the appropriate definition of kernel functions that account for signal characteristics explicitly, from correlated noise to complex-valued signals and spatio-temporal structures, just to name a few of the pressing challenges in many fields of science.

\section*{Acknowledgments}
This work was partly supported by the European Research Council (ERC) under the ERC-CoG-2014 SEDAL project (647423) and project RTI2018-096765-A-100 (MCIU/AEI/FEDER, UE). \\

\subsection*{APPENDIX A: SMOS and ENSO datasets}\label{appendixA}
We use global soil moisture maps from the ESA SMOS mission, available at \href{http://bec.icm.csic.es/}{SMOS Barcelona Expert Center (BEC)}. Since its launch in 2010, SMOS provides global maps of the Earth's surface soil moisture (top 5 cm) every 3-days with a spatial resolution of $\sim$50 km and a target accuracy of 0.04 m$^3\cdot$m$^{-3}$. We selected the first seven years of SMOS observations, after its commissioning phase (from May 2010 to May 2017), and focus on the transition of the 2016 ENSO event. As suggested in \cite{Piles19RS}, ascending and descending daily orbits were temporally averaged and 5-day bins were constructed to ensure enough coverage and smooth spatio-temporal transitions; pixels with less than $30\%$ temporal coverage and latitudes higher than $60^\circ$ were not considered. Alongside SM data, we use time series of the ENSO4 climate index from \href{https://www.knmi.nl/home}{The Royal Netherlands Meteorological Institute (KNMI)}, which is calculated daily based on on Sea Surface Temperature (SST) anomalies  averaged across the central equatorial Pacific Ocean (5N-5S, 160E-150W). ENSO4 time series were temporally averaged into 5-day bins for this study.  

\subsection*{APENDIX B: Spatialization of the causality index}

\begin{figure}[hbt!]
\centerline{\includegraphics[width=10cm]{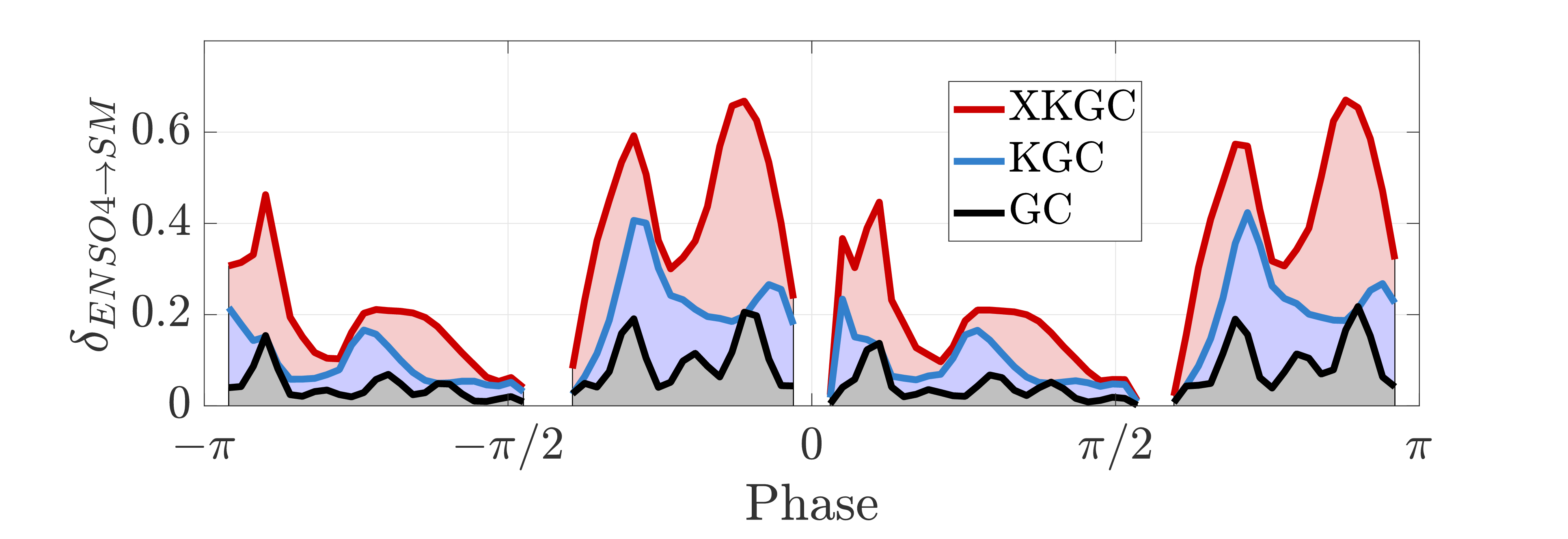}}
\vspace{-0.25cm}
\caption{Dependence of $\delta_{ENSO4 \rightarrow SM}$ with the phase. Note that all methods preserve the non-causal phase dependence. The causality index-phase relation is cyclic with $\pi$ period. \label{fig:delta_phase}}
\end{figure}

The ROCK-PCA method extracts time series and spatial components in the complex domain, which can be connected in the phase space, i.e. the phase of the spatial component is the phase of the time series for each pixel. Hence, searching for the dependence of $\delta$ with the phase (mixture of real and imaginary time series component as we show in figure~\ref{fig:delta_phase}), we can spatialize our results. This phase dependence is the temporal mean for each moving window $\delta$ estimation. 

\begin{figure}[hbt!]
\centerline{\includegraphics[width=8.5cm]{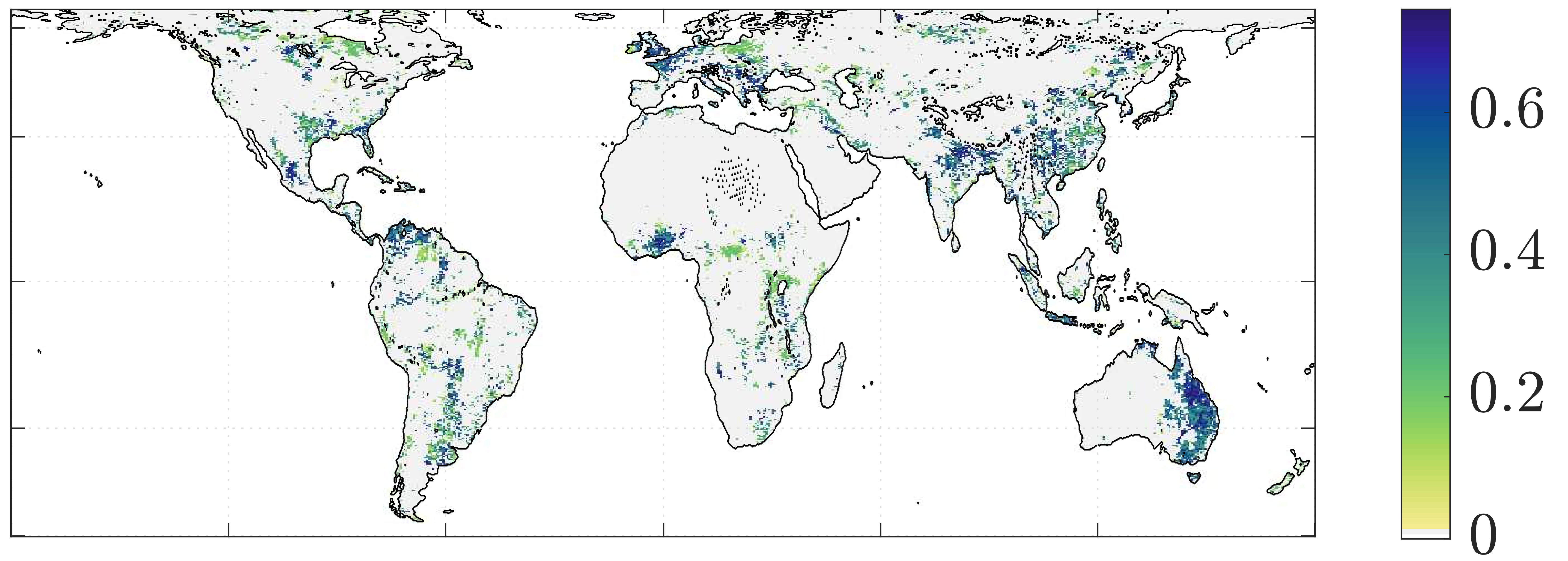}}
\vspace{-0.25cm}
\caption{Global distribution of the ENSO-SM causality index $\delta_{ENSO4 \rightarrow SM}$. A greater $\delta$ indicates a higher percentage of SM interannual variability can be explained by ENSO.
\label{fig:delta_map}}
\end{figure}

The relation between $\delta$ and each pixel using the transformation of the curve of figure~\ref{fig:delta_phase} allows us to obtain the spatial map of $\delta$, which is shown in Fig. ~\ref{fig:delta_map} for the XKGC method. Regions with causal representation but without SM variability (e.g. desserts) were masked in Fig.~\ref{fig:delta_map}. Interestingly, regions with clearly differentiated $\delta$ levels emerge. This map can be interpreted as a forecast skill, where differences in skill are caused by the different underlying mechanisms involved in the ENSO and SM relation in each region. Most of causal regions are represented over the tropics as we expect, but other supra tropical regions, as North-West Europe, emerge. 

\bibliography{references,extra}

\end{document}

%% file: definitions.tex
\usepackage[colorlinks=true,citecolor=blue]{hyperref}
\usepackage{color}

\newcommand{\real}{\mathbb{R}}
\newcommand{\x}{{\bf x}}
\newcommand{\y}{{\bf y}}
\newcommand{\z}{{\bf z}}

\newcommand{\remove}[1]{}

\newcommand{\mcalX}{\mathcal{X}}

\newcommand{\mcalH}{\mathcal{H}}

\newcommand{\balpha}{\boldsymbol{\mathit{\alpha}}}

\newcommand{\bphi}{\boldsymbol{\mathit{\phi}}}

\newcommand{\bpsi}{\boldsymbol{\mathit{\psi}}}

\usepackage{graphicx}
\usepackage{dcolumn}
\usepackage{bm}

\usepackage{graphics}
\usepackage{pgf}
\usepackage{tikz}
\usetikzlibrary{arrows,automata,positioning}
\usetikzlibrary{mindmap,trees}

\usetikzlibrary{shapes,arrows,positioning} 

\tikzset{
    >=stealth',
    punkt/.style={
           rectangle,
           rounded corners,
           draw=black, very thick,
           text width=6.5em,
           minimum height=2em,
           text centered},
    pil/.style={
           ->,
           thick,
           shorten <=2pt,
           shorten >=2pt,}
}

\usepackage{array}
\newcolumntype{P}[1]{>{\centering\arraybackslash}p{#1}}